\documentclass[a4paper]{jpconf}
\usepackage{iopams}
\usepackage{graphicx}
\usepackage{slashed} 
\usepackage{cite}
\usepackage{amsmath}
\newcommand{\chg}{\ensuremath{e}}
\newcommand{\mgm}{\ensuremath{\mu}}
\newcommand{\elm}{\ensuremath{\epsilon}}
\newcommand{\anm}{\ensuremath{a}}

\begin{document}
\title{Electromagnetic interactions of neutrinos in processes of low-energy elastic neutrino-electron scattering}

\author{Konstantin A Kouzakov$^{1}$ and Alexander I Studenikin$^{2,3}$}

\address{$^1$ Department of Nuclear Physics and Quantum Theory of Collisions, Faculty of Physics, Lomonosov Moscow State University, Moscow 119991, Russia}
\address{$^{2}$ Department of Theoretical Physics, Faculty of Physics, Lomonosov Moscow State University, Moscow 119991,
Russia\\
$^{3}$ Joint Institute for Nuclear Research, Dubna 141980, Moscow
Region, Russia}
\ead{kouzakov@gmail.com}
\begin{abstract}
Electron energy and angular distributions in the process of low-energy elastic neutrino-electron scattering are treated in the free-electron approximation. The effects of the millicharges, magnetic, electric, and anapole moments of massive neutrinos along with the flavor change of neutrinos traveling from the source to the detector are taken into account under the assumption of three-neutrino mixing. The footprints of neutrino electromagnetic interactions in the electron energy and angle distributions are discussed.
\end{abstract}
\section{Introduction}
\label{intro}
In the standard model neutrinos are massless left-handed fermions which very weakly interact with matter via exchange of the $W^\pm$ and $Z^0$ bosons. The development of our knowledge about neutrino masses and mixing~\cite{Bilenky:2010zza,Xing:2011zza,King:2015review} provides a basis for exploring neutrino properties and interactions beyond the standard model (BSM). In this respect, the study of nonvanishing electromagnetic characteristics of massive neutrinos is of particular interest~\cite{Broggini:2012df,Giunti_RMP2015,giunti16}. It can help not only to shed light on whether neutrinos are Dirac or Majorana particles, but also to constrain the existing BSM theories and/or to hint at new physics.

The possible electromagnetic properties of massive neutrinos include the electric charge (millicharge), the charge radius, the dipole magnetic and electric moments, and the anapole moment. Their effects can be searched in astrophysical environments, where neutrinos propagate in strong magnetic fields and dense matter~\cite{Raffelt:1996wa}, and in laboratory measurements of neutrinos from various sources. In the latter case, a very sensitive and widely used method is provided by the direct measurement of low-energy elastic (anti)neutrino-electron scattering in reactor, accelerator, and solar experiments. A typical strategy of such experiments consists in determining deviations of the energy spectrum of recoil electrons from the prediction of the standard model of the electroweak interaction. On the other hand, one can study the angular distribution of recoil electrons rather than their energy spectrum. The present work aims at pointing out the effects of neutrino electromagnetic interactions both on the energy and on the angular distributions of recoil electrons in the scattering experiments.  


The paper is organized as follows. Sec.~\ref{nu_e-m} gives a brief account of neutrino electromagnetic properties. In Sec.~\ref{free-electron}, general formulas for the energy and angular differential cross sections are presented. 
Then, in Sec.~\ref{rec_el}, marked features of the energy and angulaer distributions of recoil electrons are discussed. The conclusions are drawn in Sec.~\ref{concl}.

\section{Electromagnetic properties of massive neutrinos}
\label{nu_e-m}
The effective electromagnetic interaction Hamiltonian for massive neutrino fields can be presented as
\begin{equation}
{H}_{EM} = \sum_{j,k} \overline{\nu}_{j} \Lambda^{jk}_{\mu}
\nu_{k} A^{\mu} , \label{C033}
\end{equation}
where possible transitions between different massive neutrinos are also taken into account. The effective
electromagnetic vertex in momentum-space representation depends
only on the four-momentum $q=p_j-p_k$ transferred to the photon. At low neutrino energies it can be expressed as follows:
\begin{equation}
\Lambda^{jk}_{\mu}(q) = \left( \gamma_{\mu} - \frac{q_{\mu}
\slashed{q}}{q^2} \right) \left[ e_{jk} +\frac{q^2}{6} \left(\langle r^2\rangle_{jk}
+6\gamma_{5}a_{jk}\right) \right]  + \sigma_{\mu\nu} q^{\nu} \left[ \mu_{jk} +
i \gamma_{5} \epsilon_{jk} \right], \label{C043}
\end{equation}
where $\sigma_{\mu\nu}=(\gamma_\mu\gamma_\nu-\gamma_\nu\gamma_\mu)/2$, and $\chg_{jk}$, $\mgm_{jk}$, $\elm_{jk}$, and $\anm_{jk}$ are,
respectively, the neutrino millicharge, magnetic moment, electric
moment, and anapole moment of diagonal ($j=k$) and transition
($j{\neq}k$) types. 

Since a Majorana field has half the degrees of freedom of a Dirac field, its electromagnetic properties are also reduced, namely in the Majorana case the charge, magnetic and electric form-factor matrices in the mass basis are antisymmetric and the anapole form-factor matrix is symmetric. Therefore, a Majorana neutrino does not have diagonal charge and dipole magnetic and electric moments. It can only have a diagonal anapole moment. At the same time, Majorana neutrinos can have as many transition moments as Dirac neutrinos.

\section{Differential cross sections}
\label{free-electron}
We consider the process where an ultrarelativistic Dirac neutrino with flavor $\ell$ and energy $E_\nu$ originates from a source and elastically scatters on an electron in a detector at energy transfer~$T$. If the neutrino is born in the source in the flavor state $|\nu_\ell\rangle$, then its state in the detector is
\begin{equation}
\label{nu_state}|\nu_\ell(L)\rangle=\sum_{k=1}^3U^*_{\ell k}e^{-i\frac{m_k^2}{2E_\nu}L}|\nu_k\rangle,
\end{equation}
where $|\nu_k\rangle$ is the physical neutrino state with mass $m_k$ and $L$ is the source-detector distance. The simplest model of the electron system in the detector is a free-electron model, where it is assumed that electrons are free and at rest. This approximation is supposed to be generally applicable if the energy-transfer value $T$ is much larger than the electron binding energy in the detector (for the analysis and discussion of the effects beyond the free-electron approximation see, for instance, Refs.~\cite{Kouzakov:2010tx,Kouzakov:2011vx,Kouzakov:2011ka,Kouzakov:2011ig,Kouzakov:2014lka} and references therein). In the scattering experiments the observables are the kinetic energy $T_e$ of the recoil electron and/or its solid angle $\Omega_e$. From the energy-momentum conservation one gets
\begin{equation}
\label{kinemat}
T_e=T, \qquad \cos\theta_e=(1+\xi)\sqrt{\frac{T}{T+2m_e}},
\end{equation}
where $\theta_e$ is the angle of the recoil electron with respect to the neutrino beam and $\xi=m_e/E_\nu$. The cross section, which is differential with respect to the electron kinetic energy $T_e$, is given by~\cite{Kouzakov:2017prd}
\begin{equation}
\label{cr_sec_FE_1}
\frac{d\sigma^{\rm FE}}{dT_e}=\frac{d\sigma_{(w,Q)}^{\rm FE}}{dT_e}+\frac{d\sigma_{(\mu)}^{\rm FE}}{dT_e},
\end{equation}
with
\begin{eqnarray}
\frac{d\sigma_{(w,Q)}^{\rm FE}}{dT_e}&=&\frac{G_F^2m_e}{2\pi}\left\{A_\ell+B_\ell+2\Re(C_\ell)+[A_\ell+B_\ell-2\Re(C_\ell)]\left(1-\frac{T_e}{E_\nu}\right)^2
+(B_\ell-A_\ell)\frac{T_em_e}{E_\nu^2}\right\},\nonumber\\
\\
\frac{d\sigma_{(\mu)}^{\rm FE}}{dT_e}&=&\frac{\pi\alpha^2}{m_e^2}\,\left|\mu_\ell^{eff}\right|^2\left(\frac{1}{T_e}-\frac{1}{E_\nu}\right),
\end{eqnarray}
where
\begin{eqnarray}
A_\ell&=&\sum_{j,k,k'=1}^3U^*_{\ell k}U_{\ell k'}e^{-i\frac{\delta m_{kk'}^2}{2E_\nu}L}
\left[(g_V')_{jk}+\tilde{Q}_{jk}\right]\left[(g_V')_{jk'}^*+\tilde{Q}_{jk'}^*\right],\label{C1}\\
B_\ell&=&\sum_{j,k,k'=1}^3U^*_{\ell k}U_{\ell k'}e^{-i\frac{\delta m_{kk'}^2}{2E_\nu}L}
(g_A')_{jk}(g_A')_{jk'}^*,\label{C2}\\
C_\ell&=&\sum_{j,k,k'=1}^3U^*_{\ell k}U_{\ell k'}e^{-i\frac{\delta m_{kk'}^2}{2E_\nu}L}
\left[(g_V')_{jk}+\tilde{Q}_{jk}\right](g_A')_{jk'}^*,\label{C3}\\
\left|\mu_\ell^{eff}\right|^2&=&\sum_{j=1}^3\left|\sum_{k=1}^3U^*_{\ell k}e^{-i\frac{m_{k}^2}{2E_\nu}L}(\mu_\nu)_{jk}\right|^2.
\end{eqnarray}
Here $\delta m_{kk'}^2=m_k^2-m_{k'}^2$,
$$
(g_V')_{jk}=\delta_{jk}g_V+U^*_{ej}U_{ek}, \qquad (g_A')_{jk}=\delta_{jk}g_A+U^*_{ej}U_{ek},
$$
with $g_V=2\sin^2\theta_W-1/2$ and $g_A=-1/2$, and
$$
\tilde{Q}_{jk}=-\frac{\sqrt{2}\pi\alpha}{G_F}\left[\frac{(e_\nu)_{jk}}{m_eT_e}-\frac{1}{3}\langle r_\nu^2\rangle_{jk}\right].
$$
The quantities $(e_\nu)_{jk}$ (measured in units of $e$), $\langle r_\nu^2\rangle_{jk}$, and $(\mu_\nu)_{jk}$ (measured in units of $\mu_B$) are defined as follows:
\begin{equation}
(e_\nu)_{jk}=e_{jk}, \qquad \langle r_\nu^2\rangle_{jk}=\langle r^2\rangle_{jk}-6a_{jk}, \qquad  (\mu_\nu)_{jk}=\mu_{jk}-i\epsilon_{jk}. 
\end{equation}

The angular differential cross section readily derives from the energy differential cross section~(\ref{cr_sec_FE_1}), using the relations~(\ref{kinemat}). One thus gets 
\begin{equation}
\label{cr_sec_FE_2}
\frac{d\sigma^{\rm FE}}{d\Omega_e}=\frac{d\sigma_{(w,Q)}^{\rm FE}}{d\Omega_e}+\frac{d\sigma_{(\mu)}^{\rm FE}}{d\Omega_e},
\end{equation}
where  
\begin{eqnarray}
\frac{d\sigma_{(w,Q)}^{\rm FE}}{d\Omega_e}&=&\frac{G_F^2m_e^2}{\pi^2}\frac{(1+\xi)^2\cos\theta_e}{\left[(2+\xi)\xi+\sin^2\theta_e\right]^2}\Bigg\{A_\ell+B_\ell+2\Re(C_\ell)
\nonumber\\
&{}&
+[A_\ell+B_\ell-2\Re(C_\ell)]\left[\frac{\xi^2+(1+2\xi)\sin^2\theta_e}{(2+\xi)\xi+\sin^2\theta_e}\right]^2+\frac{2(B_\ell-A_\ell)\xi^2\cos^2\theta_e}{(2+\xi)\xi+\sin^2\theta_e}\Bigg\},
\end{eqnarray}
%
%
\begin{equation}
\frac{d\sigma_{(\mu)}^{\rm FE}}{d\Omega_e}=\frac{\alpha^2}{m_e^2}\,\left|\mu_\ell^{eff}\right|^2\frac{(1+\xi)^2\left[\xi^2+(1+2\xi)\sin^2\theta_e\right]}{\cos\theta_e\left[(2+\xi)\xi+\sin^2\theta_e\right]^2}.
\end{equation}
%

In the case of Dirac antineutrinos, one must make the following substitutions in the above formulas: $\Re(C_\ell)\to-\Re(C_\ell)$, $U_{\ell k}\to U_{\ell k}^*$, $(g_V')_{jk}\to-(g_V')_{jk}^*$, $(g_A')_{jk}\to-(g_A')_{jk}^*$, $(e_{\nu})_{jk}\to(e_{\bar{\nu}})_{jk}=-e_{kj},$ and
$$
\langle r_{\nu}^2\rangle_{jk}\to\langle r_{\bar{\nu}}^2\rangle_{jk}=-\langle r^2\rangle_{kj}+6a_{kj}, \qquad  (\mu_{\nu})_{jk}\to(\mu_{\bar\nu})_{jk}=-\mu_{kj}-i\epsilon_{kj}.
$$
\section{Energy and angular distributions of recoil electrons}
\label{rec_el}
The energy distribution of recoil electrons $\mathcal{N}_{e^-}(T_e)$ is determined by the differential cross section~(\ref{cr_sec_FE_1}). The kinetic energy of recoil electrons ranges from 0 to $T_e^{\rm max}$, where
$$
T_e^{\rm max}=\frac{E_\nu}{1+\frac{1}{2}\,\xi}.
$$
When $T_e\to0$ the contributions to the recoil-electron spectrum due to weak, millicharge, and magnetic scattering channels exhibit qualitatively different dependencies on $T_e$, namely 
\begin{equation}
\mathcal{N}_{e^-}^{(w,Q)}(T_e\to0)\propto\left\{
\begin{array}{cr} \displaystyle {\rm const} & (e_\nu=0),\\
\displaystyle\frac{1}{T_e^2} & (e_\nu\neq0),
\end{array}\right. \qquad \mathcal{N}_{e^-}^{(\mu)}(T_e\to0)\propto\frac{1}{T_e}.
\end{equation}

The angular distribution of recoil electrons $\mathcal{P}_{e^-}(\Omega_e)$ is confined within the forward hemisphere, i.e., $0\leq\theta_e\leq\pi/2$. According to Eq.~(\ref{kinemat}), the $\theta_e=0$ case amounts to $T_e=T_e^{\rm max}$, and when $\theta_e\to\pi/2$ the electron kinetic energy is $T_e\to0$. The weak-interaction, millicharge, and magnetic components of the electron angular distribution in the vicinity of $\theta_e=\pi/2$ behave as follows: 
\begin{equation}
\mathcal{P}_{e^-}^{(w,Q)}(\theta_e\to\pi/2)\propto\left\{
\begin{array}{cr} \displaystyle \cos\theta_e & (e_\nu=0),\\
\displaystyle\frac{1}{\cos^3\theta_e} & (e_\nu\neq0),
\end{array}\right. \qquad \mathcal{P}_{e^-}^{(\mu)}(\theta_e\to\pi/2)\propto\frac{1}{\cos\theta_e}.
\end{equation}
It can be seen that the contribution to the electron angular distribution associated with the weak interaction vanishes at $\theta_e=\pi/2$. At the same time, the $e_\nu$ and $\mu_\nu$ contributions to the electron angular distribution  infinitely grow in the limit $\theta_e\to\pi/2$. This means that the closer the electron angle to $\pi/2$ the higher sensitivity to neutrino millicharges and magnetic moments one might expect when measuring the electron angular distribution.

\section{Summary and concluding remarks}
\label{concl}
We have considered theoretically the low-energy elastic neutrino-electron collision, taking into account electromagnetic interactions of massive neutrinos and the flavor change of the neutrino traveling from the source to the detector in the framework of three-neutrino mixing. The recoil-electron energy and angular distributions in the free-electron approximation have been brought into focus. The manifestations of the neutrino millicharges and magnetic moments in these distributions have been discussed.

Some comments should be made about the applicability of the free-electron approximation. When the energy-transfer value $T$ is comparable to the electron binding energy, this approximation becomes not generally valid anymore. In particular, for atomic electrons it was found that with decreasing the $T$ value the contribution to the cross section associated with the neutrino millicharge exhibits strong enhancement as compared to the free-electron case~\cite{Chen:2014q_nu}. This is the so-called atomic ionization effect, which is observed for ultrarelativistic charged projectiles and which can be estimated within the equivalent photon approximation. At the same time, if the neutrino millicharges are zero, i.e., $e_{jk}=0$, the energy differential cross section for neutrino scattering on atomic electrons is well approximated by the stepping formula
\begin{equation}
\label{cr_sec_step}
\frac{d\sigma}{dT}=\frac{d\sigma^{\rm FE}}{dT}\sum_{\beta}n_\beta\theta(T-\varepsilon_\beta),
\end{equation}
where $n_\beta$ and $\varepsilon_\beta$ are the number and binding energy of electrons in the (sub)shell $\beta$. The stepping approximation was first introduced in Ref.~\cite{Kopeikin:1997step} on the basis of numerical calculations for the case of an iodine atomic target, and later it was supported by a general theoretical analysis~\cite{Kouzakov:2011vx,Kouzakov:2014lka}. Notable deviations of the weak and magnetic cross sections from the stepping formula~(\ref{cr_sec_step}) are found only close to the ionization threshold~\cite{Chen:2013lba,Kouzakov:2014pepan_lett}, where the cross-section values decrease relative to the free-electron approximation. The latter behavior is attributed to the effects of electron-electron correlations in atoms~\cite{Kouzakov:2014lka}.

The impact of electron-binding effects on the angular differential cross section $d\sigma/d\Omega_e$ is, in general, more involved. In particular, after being knocked by the neutrino the atomic electron can recoil from the atomic nucleus. This leads to a nonzero electron angular distribution in the backward hemisphere, what is kinematically forbidden in the free-electron model.

%
%
\ack
%
This work was supported by the Russian Foundation for Basic Research under grants
No.~16-02-01023\,A and No.~17-52-53133\,GFEN\_a.
\section*{References}
\bibliography{emp_prd}

\providecommand{\newblock}{}
\begin{thebibliography}{10}
\expandafter\ifx\csname url\endcsname\relax
  \def\url#1{{\tt #1}}\fi
\expandafter\ifx\csname urlprefix\endcsname\relax\def\urlprefix{URL }\fi
\providecommand{\eprint}[2][]{\url{#2}}

\bibitem{Bilenky:2010zza}
Bilenky S 2010 {\em {Introduction to the Physics of Massive and Mixed
  Neutrinos}\/} (New York: Springer)

\bibitem{Xing:2011zza}
Xing Z~z and Zhou S 2011 {\em {Neutrinos in Particle Physics, Astronomy and
  Cosmology}\/} (Zhejiang: Zhejiang University Press)

\bibitem{King:2015review}
King S~F 2015 {\em J. Phys. G: Nucl. Part. Phys.\/} {\bf 42} 123001

\bibitem{Broggini:2012df}
Broggini C, Giunti C and Studenikin A 2012 {\em Adv. High Energy Phys.\/} {\bf
  2012} 459526

\bibitem{Giunti_RMP2015}
Giunti C and Studenikin A 2015 {\em Rev. Mod. Phys.\/} {\bf 87} 531

\bibitem{giunti16}
Giunti C, Kouzakov K~A, Li Y~F, Lokhov A~V, Studenikin A~I and Zhou S 2016 {\em
  Ann. Phys. (Berlin)\/} {\bf 528} 198

\bibitem{Raffelt:1996wa}
Raffelt G 1996 {\em {Stars as Laboratories for Fundamental Physics: The
  Astrophysics of Neutrinos, Axions, and Other Weakly Interacting Particles}\/}
  (Chicago: University of Chicago Press)

\bibitem{Kouzakov:2010tx}
Kouzakov K~A and Studenikin A~I 2011 {\em Phys. Lett. B\/} {\bf 696} 252

\bibitem{Kouzakov:2011vx}
Kouzakov K~A, Studenikin A~I and Voloshin M~B 2011 {\em Phys. Rev. D\/} {\bf
  83} 113001

\bibitem{Kouzakov:2011ka}
Kouzakov K~A, Studenikin A~I and Voloshin M~B 2011 {\em JETP Lett.\/} {\bf 93}
  699

\bibitem{Kouzakov:2011ig}
Kouzakov K~A and Studenikin A~I 2011 {\em Nucl. Phys. Proc. Suppl.\/} {\bf 217}
  353

\bibitem{Kouzakov:2014lka}
Kouzakov K~A and Studenikin A~I 2014 {\em Adv. High Energy Phys.\/} {\bf 2014}
  569409

\bibitem{Kouzakov:2017prd}
Kouzakov K~A and Studenikin A~I 2017 {\em Phys. Rev. D\/} {\bf 95} 055013

\bibitem{Chen:2014q_nu}
Chen J~W, Chi H~C, Li H~B, Liu C~P, Singh L, Wong H~T, Wu C~L and Wu C~P 2014
  {\em Phys. Rev. D\/} {\bf 90} 011301

\bibitem{Kopeikin:1997step}
Kopeikin V~I, Mikaelyan L~A, Sinev V~V and Fayans S~A 1997 {\em Phys. At.
  Nucl.\/} {\bf 60} 1859

\bibitem{Chen:2013lba}
Chen J~W, Chi H~C, Huang K~N, Liu C~P, Shiao H~T, Singh L, Wong H~T, Wu C~L and
  Wu C~P 2014 {\em Phys. Lett. B\/} {\bf 731} 159

\bibitem{Kouzakov:2014pepan_lett}
Kouzakov K~A and Studenikin A~I 2014 {\em Phys. Part. Nucl. Lett.\/} {\bf 2014}
  458

\end{thebibliography}
\end{document}